\documentclass[pra,aps,twocolumn,showpacs,epsfig]{revtex4}
\newcommand{\beq}{\begin{equation}}
\newcommand{\eeq}{\end{equation}}
\newcommand{\beqa}{\begin{eqnarray}}
\newcommand{\eeqa}{\end{eqnarray}}
\newcommand{\ba}{\begin{array}}
\newcommand{\ea}{\end{array}}
\usepackage{color}
\usepackage[dvips]{epsfig}

\begin{document}

\title{Coherence and entanglement in \\
the ground-state of a bosonic Josephson junction:\\
from macroscopic Schr\"odinger cats to separable Fock states}

\author{G.Mazzarella$^{1}$, L. Salasnich$^{1,2}$, A. Parola $^{3}$,
and F. Toigo$^{1}$}

\affiliation{$^{1}$Dipartimento di Fisica ``Galileo Galilei''
and Consorzio Nazionale Interuniversitario per la Scienze Fisiche della
Materia (CNISM), Universit$\grave{a}$ di Padova,
Via Marzolo 8, I-35122 Padova, Italy \\
$^{2}$Istituto Nazionale di Ottica (INO) del
Consiglio Nazionale delle Ricerche (CNR),
via G. Sansone 1, 50019 Sesto Fiorentino, Italy\\
$^{3}$Dipartimento di Fisica e Matematica and CNISM,\\
Universit$\grave{a}$ dell'Insubria, Via Valleggio 11, I-22100 Como, Italy}

\date{\today}

\begin{abstract}
We consider a bosonic Josephson junction made of $N$ ultracold and dilute
atoms confined by a quasi one-dimensional double-well potential within the
two-site Bose-Hubbard model framework. The behaviour of the system is
investigated at zero temperature by varying the inter-atomic 
interaction from the strongly
attractive regime to the repulsive one. We show that the
ground-state exhibits a crossover from a macroscopic Schr\"odinger-cat
state to a separable Fock state through an atomic coherent regime. 
By diagonalizing the Bose-Hubbard Hamiltonian we characterize the emergence of
the mascroscopic cat states by calculating the Fisher
information $F$, the coherence by means of  the visibility $\alpha$ of the 
interference fringes in the momentum distribution, and the quantum correlations
by using the entanglement entropy $S$. 
Both Fisher information and visibility are shown to be related to 
the ground state energy 
by employing the Hellmann-Feynman theorem. This result, 
together with a perturbative calculation of the ground-state energy, 
makes possible to obtain
simple analytical formulas for $F$ and $\alpha$ over a range of interactions, 
in excellent agreement with the exact diagonalization of the 
Bose-Hubbard Hamiltonian. In the attractive regime the 
entanglement entropy attains values very close to its upper limit for 
a specific interaction strength
lying in the region where coherence is lost and self trapping sets in.  
\end{abstract}

\pacs{03.75.Ss,03.75.Hh,64.75.+g}

\maketitle
%%%%%%%%%%%%%%%%%%

\section{Introduction}

Ultracold and dilute alkali-metal atoms confined in a
quasi one-dimensional (1D) double-well potential \cite{oliver}
are the ideal system to study the Josephson effect
\cite{book-barone} and, more generally, the formation
of macroscopic coherent states \cite{smerzi,stringa,anglin,mahmud,anna}
and macroscopic Schr\"odinger-cat states \cite{cirac,dalvit,huang,carr,brand}.
The observed coherent dynamics of the atomic cloud
in the double-well potential (bosonic Josephson junction)
\cite{oliver,exp-bec} is efficiently described by Josephson equations
\cite{smerzi}, and their extensions \cite{ajj1,ajj2,sb}.
The Josephson equations are valid within the weak
inter-atomic interaction regime, where semiclassical approximations can
be performed and the system considered is in a coherent state \cite{leggett}.
When the bosons are repulsively interacting, the crossover from a coherent
state (superfluid-like regime) to a separable Fock state 
(Mott-like regime) takes place 
by increasing the coupling strength between the atoms
\cite{stringa,anglin,mahmud,anna}. This crossover could be detected by
measuring the visibility $\alpha$ of the interference fringes
in the momentum distribution of the bosonic cloud \cite{stringa,anna}.
In the case of attractive coupling, the Josephson equations predict  the
spontaneous symmetry-breaking above a critical
strength \cite{smerzi,sb}, while the two-site Bose-Hubbard model
{\bf predicts} the formation of a macroscopic Schr\"odinger-cat state
\cite{cirac,dalvit,huang,carr,brand}. Finally, when the attraction between
the bosons is sufficiently strong the collapse of the cloud should
occur \cite{sb,io-e-boris}.

The goal of the present paper is to investigate the coherence and entanglement
of the $N$ boson ground-state, with particular attention to the
formation of macroscopic Schr\"odinger-cat states, by using as
theoretical tool the two-site Bose-Hubbard (BH) Hamiltonian. Both the
experimental implementation of two-mode quantum entangled states
and the two-site BH model are the ideal instruments to address 
the problem of the
precision of the difference number and phase measurements
\cite{oberthalernature,oberthalerpra10}.
In our work, we diagonalize the two-site BH Hamiltonian and evaluate 
the coherence visibility $\alpha$, 
the Fisher information $F$, and the entanglement entropy $S$ 
of the ground state. 
We calculate the Fisher information and the coherence visibility
by employing the Hellmann-Feynman theorem (HFT) 
(see, for example, Ref. \cite{cohen}) as
well. By jointly using this theorem and a perturbative approach for calculating
the ground-state energy, we obtain simple analytical formulas for $F$
and $\alpha$ over range of interactions where the perturbative 
calculation is applicable. 
This represents one of the main novelties of the present paper. 
It is quite interesting to compare the
results obtained by applying the HFT  and the ones deriving from the
diagonalization of the BH Hamiltonian. This comparison shows a good
agreement between the predictions of the two approaches, 
in particular for the Fisher information. 

The aforementioned quantities, i.e. the Fisher information, 
the coherence visibility and the
entanglement entropy are analyzed by widely exploring the atom-atom 
interaction range, i.e. from strongly
attractive to strongly repulsive interactions. This study
is - in particular, the analysis of the entanglement entropy -  
one of the crucial points of our
paper. From the above analysis we find that the presence of a 
macroscopic Schr\"odinger-cat
state \cite{pezze} in the double-well, which can be
detected by repeated measurements of the distribution of particles
in the two wells \cite{cirac,dalvit}, corresponds to the maximum of 
the Fisher information $F$. The
maximum of the entanglement entropy $S$ is obtained for attractive 
interactions close to the onset
of coherence loss, which corresponds to the crossover to the 
self-trapped state. This maximum value is very close to its 
maximum achievable value, i.e. $\log_{2}(N+1)$, and greater than 
the one of a macroscopic cat state.
For sufficiently strong attractions the coherence visibility 
$\alpha$ is quite small but
it increases as the interaction strength becomes sufficiently weak.
In addition, we find that, by strongly increasing the repulsive interaction,
the coherence visibility, the Fisher information $F$ and
the entanglement entropy $S$ go to zero. In this strongly-repulsive
regime the ground state becomes a separable Fock state with $N/2$ particles
in each potential well.

To make complete our analysis, we exploit the known fact 
that the Josepshson equations for 
a bosonic junction can be derived from the two-site Bose-Hubbard Hamiltonian by
using the quasi-classical coherent state; this state is characterized by
non-vanishing quantum number fluctuations. We compare the $N$-boson
ground-state energy obtained by diagonalizing the BH Hamiltonian with the
expectation value of the BH Hamiltonian evaluated with respect to 
the quasi-classical coherent state. In this way, we show the link 
between the quantum description and the quasi-classical 
one of the bosonic Josephson junction.

\section{The model Hamiltonian}

We consider a dilute and ultracold atomic gas made of
$N$ identical bosons of mass $m$ confined
by a trapping potential $V_{trap}({\bf r})$. This
potential can be realized by the superposition of an isotropic harmonic
confinement in the the transverse radial plane
and a double-well potential $V_{DW}(x)$ in the axial direction $x$.
Then, $V_{trap}({\bf r})$ is given by
$V_{trap}({\bf r})= V_{DW}(x)+{m\omega_{\bot}^2}(y^2+z^2)/2$,
where $\omega_{\bot}$ is the trapping frequency in the radial plane.

We suppose that the system is quasi one-dimensional (1D) due to a strong
transverse radial harmonic confinement. In particular,
the transverse energy $\hbar \omega_{\bot}$ is much larger than
the characteristic energy of bosons in the axial direction.
By assuming that the two wells are symmetric,
the microscopic dynamics of our system is described by the effective
two-sites Bose-Hubbard (BH) Hamiltonian \cite{ajj2}
\beq
\hat{H} = -J\big(\hat{a}^{\dagger}_L\hat{a}_R
+\hat{a}^{\dagger}_R\hat{a}_L\big)
+\frac{U}{2}\big( \hat{n}_L (\hat{n}_L -1)
+ \hat{n}_R (\hat{n}_R -1)\big) \; ,
\label{twomode}
\eeq
where $\hat{a}_{k},\hat{a}^{\dagger}_{k}$ ($k=L,R$)
are bosonic operators satisfying the algebra
$[\hat{a}_{k},\hat{a}^{\dagger}_{l}]=\delta_{kl}$;
$\hat{n}_{k}=\hat{a}^{\dagger}_{k}\hat{a}_{k}$ is the number of
particles in the $k$th well; $U$
is the boson-boson interaction amplitude; $J$ is the tunnel matrix
element between the two wells.
Note that the sign of the hopping amplitude can be changed by gauge
transformation:
in the following we will refer to $J$ as a positive quantity.
We observe that the total number operator
\beq
\hat{N}= \hat{n}_{L} + \hat{n}_{R}
\eeq
commutes with the Hamiltonian (\ref{twomode}), so that the total number of
particles is conserved.
The macroscopic parameters in the Hamiltonian (\ref{twomode})
are explicitly related to the atom-atom coupling
constant $g=4\pi\hbar^2a_s/m$ - $a_s$ being the s-wave scattering
length - and to the other microscopic parameters, i.e. the atomic mass
$m$ and the frequency $\omega_{\bot}$
of the harmonic trap (see Ref. \cite{ajj2}).
The on-site interaction amplitude $U$ is positive (negative) if $a_s$ is
positive (negative), so that it may be changed by Feshbach resonance.
The hopping amplitude $J$ is equal to
$(\epsilon_{1}-\epsilon_{0})/2$, where $\epsilon_{0}$ and
$\epsilon_{1}$ are the ground-state and first excited state energies of a 
single boson in the double-well \cite{fermidyn}.

\section{Coherent, Fock and cat states}

At fixed number $N$ of bosons, the ground-state of the system depends
on the adimensional parameter $\zeta = U/J$. Let us recall what is known
in the literature.

\begin{itemize}

\item
In the case of zero interaction ($U=0$) the ground-state
of the BH Hamiltonian with a fixed total number $N$ of bosons
is the atomic coherent state (ACS)
\beq
\label{coherent}
|ACS\rangle= {1\over \sqrt{N!}}
\left[ {1\over \sqrt{2}}({\hat a}_{L}^{\dagger} + {\hat a}_{R}^{\dagger})
\right]^N |0,0\rangle
\; ,
\eeq
where $|0,0\rangle = |0\rangle_L \otimes |0\rangle_R$, that is
the tensor product between the vacuum of the operator
${\hat a}_L$ - $ |0\rangle_L$ - and the vacuum of and ${\hat a}_R$,
$|0\rangle_R$ \cite{arecchi,diaz}.

\item
In the case of repulsive interaction ($U>0$)
the ground-state of the BH Hamiltonian can be found
into two very different regimes: the superfluid-like regime
and the Mott-like regime \cite{stringa,anglin,mahmud,anna}.
The transition from one regime to the other depends on
the parameter $\zeta = U/J >0$.
If $\zeta \ll N$, the system is in a quasiclassical superfluid
regime characterized by the
suppression of phase fluctuations: the ground-state
is close to $|ACS \rangle$. If $\zeta \gg N$, the BH Hamiltonian
(\ref{twomode}) yields the Mott-insulator like regime where number
fluctuations are frozen: the ground-state
is incoherent \cite{smerzi,stringa,anglin,mahmud,anna},
and close to the separable Fock state
\beq
\label{Fock}
|FOCK \rangle = |{N\over 2},{N\over 2}\rangle \; .
\eeq

\item
In the case of attractive interaction ($U<0$)
the ground-state of the BH Hamiltonian
can be close to the atomic coherent state $|ACS \rangle$ or
in a entangled superposition of macroscopic states,
i.e. macroscopic Schr\"{o}dinger cat states
\cite{cirac,dalvit,huang,carr,brand}.
By changing the parameter $\zeta = U/J<0$
the ground-state of the system evolves towards the following
macroscopic superposition state
\beq
\label{cat}
|CAT \rangle = {1\over \sqrt{2}}(|N,0\rangle + |0,N\rangle ) \;
\eeq
which is the linear combination of the states with all particles
in the left or in the right well. This entangled macroscopic
superposition state is also known as ``$NOON$ state'' or ``macroscopic
cat state''.
The lower is $\zeta$ the more likely the system is expected be in the
macroscopic Schr\"odinger-cat state \cite{cirac,dalvit,huang,anna,carr,brand}.

\end{itemize}

A very interesting issue to address is the study of the 
coherence and the entanglement in a bosonic Josephson junction
by varying the ratio $U/J$ from negative to positive values 
keeping fixed the number $N$ of atoms.
To accomplish this task  we solve numerically the following eigenproblem
\beq
{\hat H} |E_j\rangle = E_j |E_j \rangle
\eeq
for a fixed number $N$ of bosons. In this case
the Hamiltonian ${\hat H}$ can be represented by
a $(N+1)\times (N+1)$ matrix in the Fock basis $|i,N-i\rangle
=|i\rangle_L \otimes |N-i\rangle_R$
with $i=0,...,N$. For each eigenvalue $E_j$, with
$j=0,1,...,N$, the corresponding eigenstate $|E_j\rangle$ will be of the form
\beq
|E_j\rangle=\sum_{i=0}^{N}\,c_{i}^{(j)} \, |i,N-i\rangle \; .
\label{eigenstate}
\eeq

\begin{figure}[ht]
\epsfig{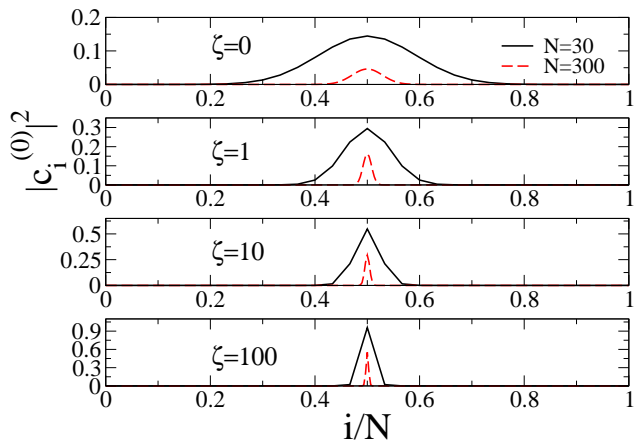}
\caption{(Color online). Coefficients $|c_{i}^{(0)}|^2$ of the
repulsive ($U>0$)
ground-state $|E_0\rangle$ (see Eq. (\ref{eigenstate}))
of the bosonic junction as a function of $i/N$.
$|c_{i,0}|^2$ are plotted for four values of the adimensional parameter
$\zeta=U/J\geq 0$. Note that $\sum_{i=0}^N|c_i^{(0)}|^2=1$ and the
total number of coefficients is $N+1$. The coefficients $|c_{i}^{(0)}|^2$ and
$i/N$ are adimensional quantities.}
\label{fig1}
\end{figure}

\begin{figure}[ht]
\epsfig{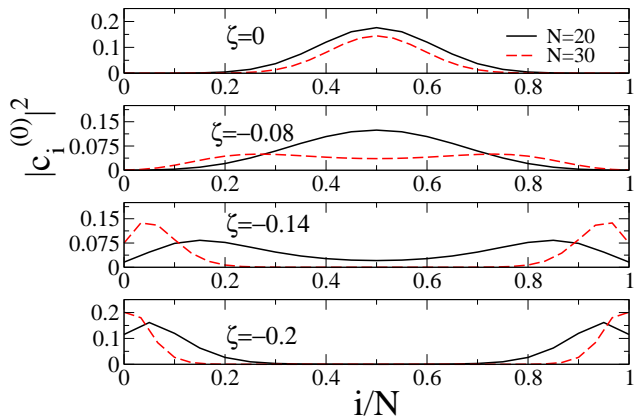}
\caption{(Color online). Coefficients $|c_{i}^{(0)}|^2$ of
the attractive ($U<0$)
ground-state $|E_0\rangle$ (see Eq. (\ref{eigenstate}))
of the bosonic junction as a function of $i/N$.
$|c_{i,0}|^2$ are plotted for four values of the adimensional parameter
$\zeta=U/J \leq 0$. The coefficients $|c_{i}^{(0)}|^2$ and
$i/N$ are adimensional quantities.}
\label{fig2}
\end{figure}

In Fig. \ref{fig1} we plot the coefficients $|c_{i}^{(0)}|^2$
of the repulsive ($U>0$) ground-state $|E_0\rangle$ for different
values of the parameter $\zeta=U/J$. In correspondence to each value of
$\zeta$, we have studied the ground-state for $N=30$ (solid line) and $N=300$
(dashed line). Let us focus on the solid line. The figure shows
that the width of the peak of $|c_{i}^{(0)}|^2$ becomes narrower
by increasing $\zeta$. The hardening of the localization corresponds
to a reduction of the coherence. This is related to the fact that 
the ground-state $|E_0\rangle$ is becoming
the separable Fock state $|FOCK\rangle$. The same kind of behavior is observed
if the number of particles $N$ is increased. From the plots reported in
Fig. \ref{fig1}, we can see that, given $\zeta$, the greater is $N$ 
the stronger is the localization.

In Fig. \ref{fig2} we plot the coefficients $|c_{i}^{(0)}|^2$
of the attractive ($U<0$) ground-state $|E_0\rangle$ for different
values of the parameter $\zeta=U/J$. In correspondence to each value of
$\zeta$, we have analyzed the ground-state for $N=20$ (solid line) and $N=30$
(dashed line). Once fixed the number of bosons $N$, we see that
by increasing $\zeta$ a crossover occurs.
It exists a value of $\zeta$, say $\zeta_{cr}$ ($\zeta_{cr} \simeq
-0.116$ for $N=20$ and $\zeta_{cr}\simeq -0.076$ for $N=30$), for which
a valley appears in the middle of $|c_{i}^{(0)}|^2$. This is a signature of
the fact that the system is losing coherence and the junction 
is approaching the self-trapping regime.
The above mentioned crossover takes place from a ground-state which has
the maximal probability at $i=N/2$ (single peak centered around $i=0$) 
for $\zeta=0$, to a ground-state which has
the same maximal probability at $i=n$ and $i=N-n$ (two separated peaks
symmetric with respect to $i=0$) with $n$ approaching $0$ as $\zeta$ 
decreases. 
Thus,  for a sufficiently small $\zeta$ 
(i.e. $\zeta<0$ and $|\zeta|$ large) the ground-state $|E_0\rangle$
has the maximal probability to be in the state in which the bosonic
population in each well fluctuates around $N/2$.
For a sufficiently large $\zeta$, instead, the ground state $|E_0\rangle$
has the maximal
probability to be in a ``macroscopic cat state'': a linear superposition
of states with macroscopically different occupations.
From the plots shown in Fig. \ref{fig2}, we can see that the larger is $N$,
the larger is the attraction for which the crossover to  the ``macroscopic cat
state'' occurs.

Finally, we can summarize that Fig. \ref{fig1} and Fig. \ref{fig2} show
the crossover from the ``macroscopic cat state''
$|CAT\rangle$ for $\zeta \ll 0$ to the "separable Fock state" 
$|FOCK\rangle$ for $\zeta \gg 0$, crossing
the atomic coherent state $|ACS\rangle$ at $\zeta=0$.

It is quite important to note that the ground-state $|E_{0} \rangle$ has
definite parity, namely the distribution of the coefficients $c_{i}^{(0)}$ is
even with respect to $i/N=1/2$. As consequence,
we have that $\langle \hat{n}_{L} \rangle=N/2=\langle \hat{n}_{R} \rangle$.

\begin{figure}[h]
\centering
\epsfig{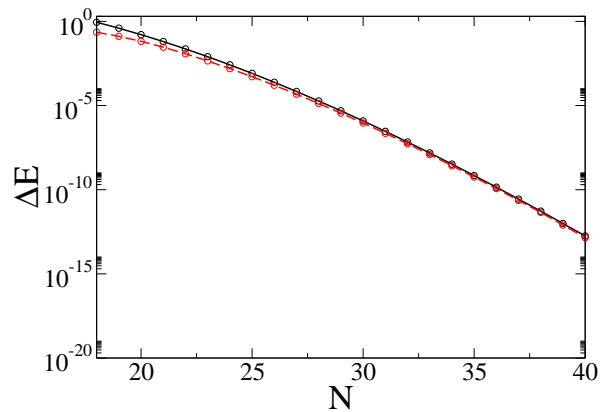}
\caption{(Color online). On the vertical axis the energetic gap $\Delta E$
at $U/J=-0.15$. On the
horizontal axis the number of bosons $N$. The solid line
is obtained with Eq. (\ref{gap}).
The circles on the dashed line are the points obtained from the
diagonalization of the Hamiltonian (\ref{twomode}). 
Note that the vertical axis is in logarithmic scale.
$\Delta E$ is in units of $J$.}
\label{gapf}
\end{figure}

To conclude this section we want to stress 
the role played by the gap, $\Delta E$, 
between the ground-state and the first excited state of the
Hamiltonian (\ref{twomode}). 
In fact, when this gap approaches zero the even and the odd states become
quasi-degenerate allowing parity symmetry breaking.
Let us focus on the attractive interactions, $U<0$.
In the classical limit, $U\to -\infty$, the ground state is doubly
degenerate comprising the
two Fock states $|N,0 \rangle$ and $|0,N \rangle$. The degeneracy
is lifted by quantum fluctuations and
a gap between the even and the odd linear combination of the two Fock
states sets in.
By standard semiclassical analysis, see Ref. \cite{landau}, we can estimate
the gap $\Delta E$ between
the two quasi-degenerate ground states in the $|U|N \rightarrow \infty$ limit:
\beq
\label{gap}
\Delta E =\frac{|U|\,\sqrt{N(N+2)}}{\int_0^1 dx
\,e^{-Nx^2[2\xi+1+2\ln x]}}\, e^{-N \xi}\;,
\eeq
where $\xi$ is a positive quantity given by
\beq
\label{csi}
\xi=\log (\frac{|U|N}{J})-1\;.
\eeq
This asymptotic expression holds for sufficiently large values of the
product $U\,N$, as
shown in Fig. \ref{gapf}, where $\Delta E$ is plotted as a function of
$N$ for $U/J=-0.15$.
We can see that the diagonalization of the Hamiltonian (\ref{twomode})
predicts a gap closing in $N$ according a law - dashed line -
which is well represented by the one given by Eq. (\ref{gap}), solid line
of  Fig. \ref{gapf}.

\section{Coherence and entanglement at zero temperature}

In this section we shall quantitatively characterize the 
coherence and the entanglement
properties of the ground-state, by devoting particular 
attention to the emergence of the
``macroscopic cat state".  Let us start with this last issue. 
To this end, we recall the definition of the quantum version of the Fisher
information $F_{QFI}$
\cite{braunstein,pezze,weiss}
\beq
\label{qfi}
F_{QFI}=(\Delta\hat{n}_{L,R})^2\; ,
\eeq
where
\beq
\label{variance}
(\Delta\hat{n}_{L,R})^2 =\langle (\hat{n}_{L}-\hat{n}_{R})^{2}\rangle-
\big(\langle\hat{n}_{L}-\hat{n}_{R}\rangle\big)^2
\;.\eeq
It is worth observing that the above definition 
of the Fisher information, that we use hereafter, only holds for pure states 
(i.e. only for not mixed states). 
The expectation values at the right hand side of Eqs. (\ref{qfi}) and Eqs.
(\ref{variance}) are taken with respect to the ground state $|
E_0\rangle$ which, having definite parity, satisfies the condition $\langle
\hat{n}_L\rangle=\langle \hat{n}_R \rangle$. In terms of the coefficients
$c_{i}^{(0)}$, $F_{QFI}$ is given by:
\beq
\label{qfigs}
F_{QFI} =
\sum_{i=0}^{N}\big[2i-N\big]^{2}|c_{i}^{(0)}|^2 \; .
\eeq
It is convenient to normalize $F_{QFI}$ at its maximum value $N^2$ by defining
the Fisher information $F$ as
\beq
\label{fi}
F=\frac{F_{QFI}}{N^2} \; .
\eeq
\begin{figure}[ht]
\epsfig{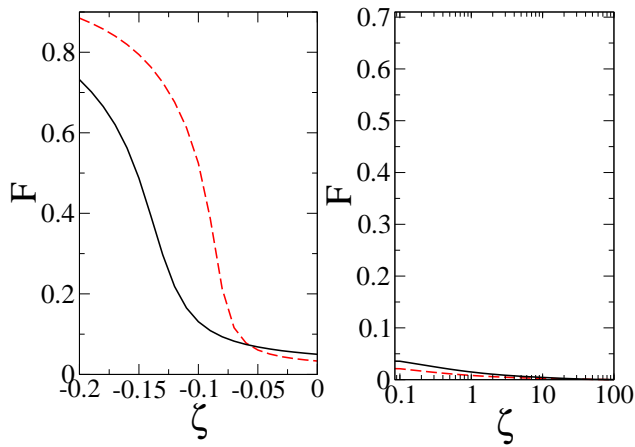}
\caption{(Color online). Fisher information $F$ of the
ground-state $|E_0\rangle$ of the bosonic junction
as a function of the parameter $\zeta = U/J$.
Left panel: attractive bosons ($U< 0$). Right panel:
repulsive bosons ($U>0$). Solid line: $N=20$. Dashed line: $N=30$.
Note that the horizontal axis of the right panel is
in logarithmic scale. The Fisher information $F$ and $\zeta$ are
adimensional quantities.}
\label{fig3}
\end{figure}

$F$ gives the width of the distribution $|c_{i}^{(0)}|^2$
centered at $i/N =1/2$. As expected, $F= 1$ holds
for the ``macroscopic cat state'' $|CAT\rangle$, and $F = 0$
for the ``separable Fock state'' $|FOCK\rangle$.

We have studied $F$ as a function of the parameter $\zeta = U/J$.
The results are shown in Fig. \ref{fig3}
for $N=20$ (solid line) and $N=30$ (dashed line).
The Fisher Information $F$ is a monotonic decreasing function of
$\zeta$ both for attractive and repulsive interactions.
On the attractive side - left panel of Fig. \ref{fig3} -
it exists a value of $\zeta$, say $\zeta_{F}$ ($\zeta_{F} \simeq -0.135$ 
for $N=20$ and
$\simeq -0.08$ for $N=30$) for which the second derivative of $F$ is zero.

It is quite interesting, now, to observe that by using the Hellmann-Feynman
theorem (HFT) - see, for example, Ref. \cite{cohen} - a relation can be 
established between the Fisher information $F$ and
the partial derivative of the ground-state energy $E_0$ with respect to $U$,
$\displaystyle{\frac{\partial E_0}{\partial U}}$. According to the HFT,
we have that
\beq
\frac{\partial E_0}{\partial U}=\langle E_0|\frac{\partial \hat{H}}
{\partial U}|E_0\rangle
\;.\eeq
The aforementioned relation is given by the following formula:
\beq
\label{fihf}
F=\frac{4}{N^{2}}\frac{\partial E_0}{\partial U}+\frac{2}{N}-1
\; .\eeq
In the limit $U/J \rightarrow -\infty$, the hopping operator in 
Hamiltonian (\ref{twomode})
can be treated within perturbation theory. To the zero order, $J=0$, the
energy $E_{0}^{(0)}$ of the ground-state of the on-site interaction operator is
$\displaystyle{\frac{UN}{2}(N-1)}$. This level is twofold degenerate,
and the two states  corresponding to this energy satisfying $\langle\hat{n}_L
\rangle=N/2=\langle\hat{n}_R\rangle$, i.e.
$\displaystyle{\frac{1}{\sqrt{2}}(|N,0\rangle+|0,N\rangle})$ and
$\displaystyle{\frac{1}{\sqrt{2}}(|N,0\rangle-|0,N\rangle})$,  have
opposite parity.

At the first non vanishing order the ground state energy is given
by \cite{note}:

\beq
\label{ae}
E_0 \simeq E_{0}^{(0)}+\frac{J^2}{U}\,\frac{N}{N-1}\;.\eeq
By substituting this energy into the expression (\ref{fihf}), we obtain
\beq
\label{fipa}
F=1-\frac{4}{N\,(N-1)\,\zeta^2}\;,\eeq
which approaches to $1$ as $\zeta \rightarrow -\infty$.

Also in the limit $U/J \rightarrow +\infty$, we can treat the hopping operator
of the Hamiltonian (\ref{twomode}) as a perturbation to the $J=0$ BH
Hamiltonian (\ref{twomode}). To the lowest order, the ground state
energy $E_{0}^{(0)}$ of the on-site interaction operator is
equal to $\displaystyle{\frac{UN}{2}(\frac{N}{2}-1)}$, which corresponds to
the separable Fock state (\ref{Fock}).
At the first non vanishing order we get
\beq
\label{re}
E_0 \simeq E_{0}^{(0)}-\frac{J^2\,N}{U}\,\big(\frac{N}{2}+1\big)\;.
\eeq
By substituting this energy into the expression (\ref{fihf}), we obtain
\beq
\label{fipr}
F=\frac{2\,(N+2)}{N\,\zeta^2}\;,\eeq
which goes to zero as $\zeta \rightarrow +\infty$.

\begin{figure}[ht]
\epsfig{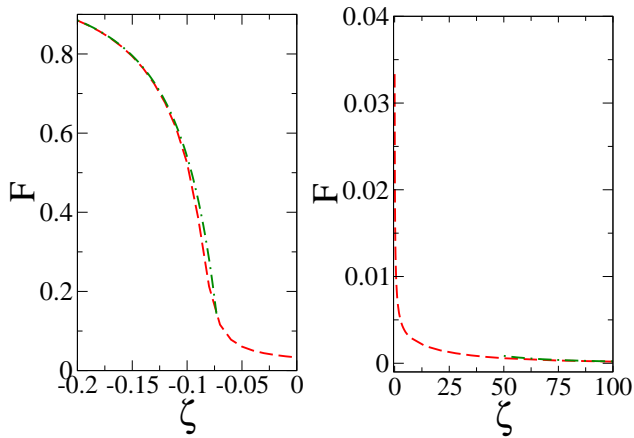}
\caption{(Color online). Fisher information $F$ of the
ground-state $|E_0\rangle$ of the bosonic junction
as a function of the parameter $\zeta = U/J$ for $N=30$.
Left panel: attractive bosons ($U<0$). Right panel:
repulsive bosons ($U>0$). Dashed line: $F$ obtained from the
diagonalization of
the Hamiltonian (\ref{twomode}). Dot-dashed line: $F$ obtained from Eq.
(\ref{fipa}) ($U<0$) and Eq. (\ref{fipr}) ($U>0$). 
The Fisher information $F$ and
$\zeta$ are adimensional quantities.}
%Note that the horizontal axis of the right panel is
%in logarithmic scale.}
\label{fig4a}
\end{figure}

In the opposite limit, $U \rightarrow 0$,
the on-site interaction term can be considered as a perturbation
in the Hamiltonian (\ref{twomode}).
For $U=0$, the ground-state energy $E_{0}^{(0)}$ of the
hopping operator is equal to $-NJ$ which corresponds to the atomic coherent
state (\ref{coherent}). To the first order in perturbation theory, we obtain
\beq
\label{cohene}
E_0 \simeq E_{0}^{(0)}+\frac{U\,N\,(N-1)}{4}\;.\eeq
By substituting this energy into the expression (\ref{fihf}), we obtain
\beq
\label{fipcoh}
F=\frac{1}{N}\;,\eeq
which tends to $0$ when {\bf $N$} tends to infinity.
In Fig. (\ref{fig4a}) we show the comparison between the Fisher
information $F$  obtained by diagonalizing the Hamiltonian (\ref{twomode}), 
dashed line,
and that one obtained by using Eq. (\ref{fipa}), attractive bosons, and
Eq. (\ref{fipr}), repulsive bosons, see the dot-dashed line.

\begin{figure}[ht]
\epsfig{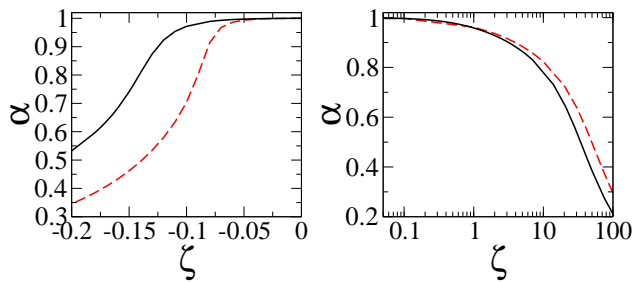}
\caption{(Color online). Coherence visibility $\alpha$ of the
ground-state $|E_0\rangle$ of the bosonic junction
as a function of the parameter $\zeta = U/J$.
Left panel: attractive bosons ($U<0$). Right panel:
repulsive bosons ($U>0$). Solid line: $N=20$. Dashed line: $N=30$.
Note that the horizontal axis of the right panel is
in logarithmic scale. The coherence visibility $\alpha$ and $\zeta$ are
adimensional quantities.}
\label{fig4}
\end{figure}

In cold atom physics it is customary to investigate the coherence
properties of condensates in terms of their momentum distribution $n(p)$,
defined as the Fourier transform of the one-body
density matrix $\rho_1(x,x')$ \cite{stringa,anglin,anna}:
\beq
\label{np}
n(p)=\int dx dx'\exp\big(-ip(x-x')\big)\,\rho_1(x,x') \; ,
\eeq
where
\beq
\rho_1(x,x')=\langle \hat{\Psi}(x)^{\dagger}\hat{\Psi}(x')\rangle \; .
\eeq
Here the operators $\hat{\Psi}(x)$ and $\hat{\Psi}(x)^{\dagger}$ -
obeying the standard bosonic commutation rules - annihilate and create,
respectively, a boson at the point $x$;  the average $\langle...\rangle$ 
is the ground-state average.
Following Refs. \cite{stringa,anglin,anna},  it is possible to show that 
the momentum distribution $n(p)$ can be written as
\beq
\label{npexp}
n(p)= n_0(p) \bigg(1 +\alpha \cos\big(pd \big) \bigg) \; .
\eeq
Here $n_0(p)$ is the momentum distribution in the fully incoherent regime
(which depends on the shape of the double-well potential $V_{DW}(x)$), and
$d$ is the distance between the two minima of $V_{DW}(x)$. $\alpha$ is
a real quantity which measures the visibility of the
interference fringes. This visibility is given by
\beq
\label{alpha}
\alpha=\frac{2\,|\langle \hat{a}^{\dagger}_L\hat{a}_R\rangle|}{N} \;
\label{visibility}
\eeq
and characterizes the degree of coherence
between the two wells.

The expectation value of the operator $\hat{a}^{\dagger}_L\hat{a}_R$
is evaluated in the ground-state $|E_0\rangle$ and the visibility
$\alpha$ is given by

\beq
\label{jpjm}
\alpha ={2\over N} \sum_{i=0}^{N}
c_{i}^{(0)} c_{i+1}^{(0)}\sqrt{(i+1)(N-i)} \; .
\eeq
We have analyzed $\alpha$ as a function of the parameter $\zeta = U/J$,
for $N=20$ (solid line) and $N=30$ (dashed line).
The results of this analysis are reported in Fig. \ref{fig4}. 
As expected, we see that the coherent state, $\zeta=0$, 
has maximum visibility, $\alpha=1$. 
On the attractive side - left panel of Fig. \ref{fig4} - it exists a
value of $\zeta$, say $\zeta_{\alpha}$,  for
which the second derivative of $\alpha$ is zero. 
We have verified that $\zeta_{\alpha} =\zeta_{F}$;
for a given $N$ such a quantity is close to the aforementioned $\zeta_{cr}$.
We observe that both for $N=20$ and for $N=30$, the coherence visibility
$\alpha$ exhibits a plateau where it is almost independent on the 
interaction strength
and very close to one. Except for this region, in correspondence of a
fixed value of $\zeta$, the greater is $N$ the smaller is $\alpha$. 
This behavior
is reversed for repulsive interactions, as shown in the right 
panel of Fig. \ref{fig4},
from which we can see that by increasing $\zeta$, $\alpha$ slowly decreases.

\begin{figure}[ht]
\epsfig{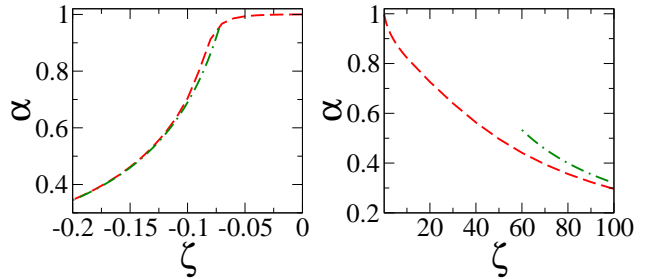}
\caption{(Color online). Coherence visibility $\alpha$ of the
ground-state $|E_0\rangle$ of the bosonic junction
as a function of the parameter $\zeta = U/J$ for $N=30$.
Left panel: attractive bosons ($U<0$). Right panel:
repulsive bosons ($U>0$). Dashed line: $\alpha$ obtained from the
diagonalization of the Hamiltonian (\ref{twomode}). Dot-dashed line:
$\alpha$ from Eq. (\ref{alphapa}) ($U <0$) and Eq. (\ref{alphapr}) ($U>0$).
The coherence visibility $\alpha$ and  $\zeta$ are adimensional quantities.}
%Note that the horizontal axis of the right panel is
%in logarithmic scale.}
\label{fig4b}
\end{figure}

\begin{figure}[ht]
\epsfig{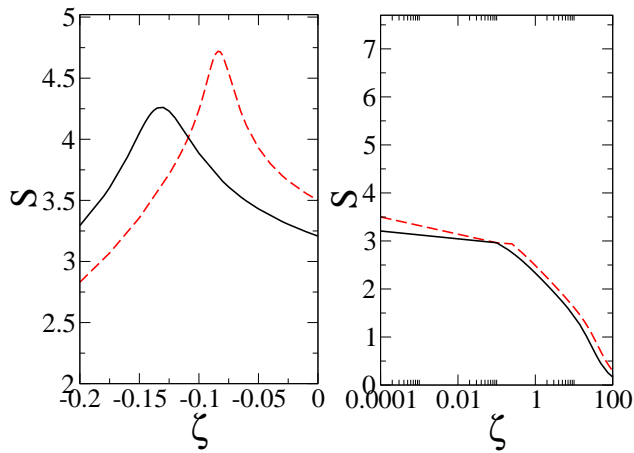}
\caption{(Color online). Entanglement entropy $S$ of the
ground-state $|E_0\rangle$ of the bosonic junction
as a function of the parameter $\zeta = U/J$.
Left panel: attractive bosons ($U<0$). Right panel:
repulsive bosons ($U>0$). Solid line: $N=20$. Dashed line: $N=30$.
Note that the horizontal axis of the right panel is
in logarithmic scale. The entanglement entropy $S$ and $\zeta$ are
adimensional quantities.}
\label{fig5}
\end{figure}
The Hellmann-Feynman theorem provides a relation between
the coherence visibility $\alpha$,
and the derivative of the ground-state energy $E_0$ with respect to $J$:
\beq
\frac{\partial E_0}{\partial J}=\langle E_0|
\frac{\partial \hat{H}}{\partial J}|E_0\rangle
\; ,\eeq
leading to
\beq
\label{jpjmhf}
\alpha =-\frac{1}{N}\frac{\partial E_0}{\partial J}\; .
\eeq
By use of the perturbative results previously obtained for the ground-state
energy $E_{0}$ - see Eqs. (\ref{ae}), (\ref{re}), (\ref{cohene}) -
we can evaluate the coherence visibility (\ref{jpjmhf})
in the three regimes $U/J \rightarrow - \infty$, $U/J \rightarrow + \infty$,
and $U
\rightarrow 0$. In the first case, Eq. (\ref{ae}) gives
\beq
\label{alphapa}
\alpha=-\frac{2}{(N-1)\zeta} \;,\eeq
which, in the limit $\zeta \rightarrow -\infty$, tends to $0$.
For $U/J \rightarrow + \infty$, Eq. (\ref{re}) provides
\beq
\label{alphapr}
\alpha=\frac{N+2}{\zeta} \;,\eeq
which tends to zero when $\zeta \rightarrow +\infty$.
Finally, at weak coupling ($U \rightarrow 0$) the perturbative result
(\ref{cohene}) gives the limiting value
\beq
\label{alphacoh}
\alpha=1 \;.\eeq
In Fig. (\ref{fig4b}) we show the comparison between the coherence 
visibility $\alpha$ obtained by diagonalizing the Hamiltonian
(\ref{twomode}), dashed line,
and the ones obtained by using Eq. (\ref{alphapa}), for attractive bosons, 
or Eq. (\ref{alphapr}), for repulsive bosons (see the dot-dashed line).

At this point, we can conclude that the joint application of the 
Hellmann-Feynman
theorem and of the perturbations theory to calculate the ground-state 
energy makes possible to obtain
very simple analytical formulas for the Fisher Information $F$, Eqs. 
(\ref{fipa}) and
(\ref{fipr}), and for the coherence visibility $\alpha$, Eqs. (\ref{alphapa}) 
and
(\ref{alphapr}), at least over a certain range of values of the interaction 
strength.

We analyze, now, the quantum entanglement pertaining to
the ground-state $|E_0\rangle$. When the system is in $|E_0\rangle$, the
density matrix $\hat{\rho}$ is
\beq
\label{dm}
\hat{\rho} =|E_0\rangle\langle E_0|
\;.\eeq
In this case, an excellent measure of the entanglement between the two wells
is provided by the entanglement entropy $S$ \cite{bwae}. This quantity
is defined as the von Neumann entropy of the reduced density 
matrix $\hat{\rho}_{L}$
defined by
\beq
\hat{\rho}_{L} =Tr_{R} \hat{\rho}
\; ,\eeq
that is a matrix obtained by partial tracing the total density matrix
(\ref{dm}) over the degrees of freedom of the right well.
The entanglement entropy $S$ is given by
\beq
\label{ee}
S=-Tr \hat{\rho_{L}} \log_{2}
\hat{\rho_{L}}=-\sum_{i=0}^{N}|c_{i}^{(0)}|^2\log_{2}|c_{i}^{(0)}|^2
\;.\eeq
For a given number of bosons $N$, the maximum value for $S$, say
$S_{(max,N)}$, is given by $\log_{2}(N+1)$. 
We have studied $S$ as a function of the parameter $\zeta = U/J$,
for $N=20$ (solid line) and $N=30$ (dashed line).
The results of this analysis - which is one of the most crucial 
studies of this paper - are reported in Fig. \ref{fig5}. 
We can see that the maximum 
entanglement is attained for a finite attractive interaction $\zeta_{S}$. 
The larger is $N$ the weaker is the attraction for which $S$ has a maximum, 
which is equal to $4.26$ when $N=20$ and to $4.72$ when $N=30$. 
These values are smaller than $S_{(max,20)}=4.39$ and $S_{(max,30)}=4.95$, 
respectively, and greater than $1$,
i.e. the entanglement entropy of a macroscopic cat state (see Tab. 1). 
The macroscopic cat state is not the maximally entangled state achievable 
in the bosonic Josephson junction. 
Moreover, we find that the interactions {\bf corresponding to} 
the entanglement entropy maxima are
$\zeta_{S} \simeq -0.13$ for $N=20$ and $\zeta_{S} \simeq -0.084$ for $N=30$.
These values are close to the $\zeta_{cr}$ - onset of the coherence
loss and the self-trapping regime within the junction dynamics - 
discussed for the attractive ground-state coefficients 
(see the discussion about Fig. \ref{fig2}).

From the right panel of Fig. \ref{fig5} we see that for repulsive
interactions,  $S$ decreases by increasing $\zeta$.

We conclude this section by summarizing - for finite even number $N$
of bosons and at zero temperature - the expansion coefficients 
$c_{i}^{(0)}$, the Fisher
information $F$, the entanglement entropy $S$, and the visibility $\alpha$ for
the atomic coherent state (\ref{coherent}), the separable 
Fock state (\ref{Fock}), 
and the "macroscopic cat state" (\ref{cat}), see Tab. 1.

\begin{widetext}
\begin{center}
\begin{tabular}{|c|c|c|c|c|}
\hline
~~~$State$~~~ & ~~~$c_{i}^{(0)}$~~~& ~~~$F$~~~ &~~~$S$~~~ & ~~~$\alpha$~~~\\
\hline
 $|ACS\rangle$ & $\displaystyle{\bigg(\frac{1}{2^{N}}\frac{N!}
{i!(N-i)!}\bigg)^{\frac{1}{2}}} $ & $\displaystyle{\frac{1}{N}}$
& $\displaystyle{\log_{2}\frac{2^{N}}{N!}+\frac{1}{2^{N-1}}
\sum_{i=2}^{N} \frac{N!}{i!(N-i)!}\log_{2}i!}$ & $1$  \\
 $|FOCK\rangle$ & 
$\delta_{i,N/2}$
& $0$ & $0$ & $0$ \\
 $|CAT\rangle$ & $\displaystyle{\frac{1}{\sqrt{2}}}
\left [ \delta_{i,0}+\delta_{i,N}\right ]$
& $1$ & $1$ & $0$\\
\hline
\end{tabular}
\end{center}
\end{widetext}
Table 1. {\small The coefficients  $c_{i}^{(0)}$, the zero-temperature Fisher
information $F$,
entanglement entropy $S$, and visibility $\alpha$ for the atomic coherent state
(\ref{coherent}), the separable Fock state (\ref{Fock}), 
and the "macroscopic cat state" (\ref{cat}) in the presence 
of a finite even number $N$ of bosons.}

\subsection{Results with two bosons}

When $N=2$, it is straightforward to calculate $E_j$ and $|E_j\rangle$.
%Let us start with repulsive bosons, i.e. $U>0$.
The ground-state
$|E_{0}\rangle$ reads
\beq
\label{gstwor}
|E_{0}\rangle=A\bigg(|2,0\rangle+\frac{\zeta+\sqrt{\zeta^2+16}}{2
\sqrt{2}}|1,1\rangle+|0,2\rangle\bigg)
\;, \eeq
where the $A$ is the constant of the normalization which is
\beq
A=\frac{2}{\sqrt{16+\zeta^2+\zeta\sqrt{\zeta^2+16}}}
\;.\eeq
The energy $E_0$ is
\beq
\label{gsetwor}
E_0=\frac{U-\sqrt{U^2+16\,J^2}}{2}
\;.\eeq
%Note that when the tunneling matrix element $J$ is zero,
%the ground-state is the separable Fock state $|1,1\rangle$, the energy of
%which is taken to be zero.
The Fisher information (\ref{fi}) calculated with respect to the state
(\ref{gstwor}) is given by the following expression
\beq
\label{fitwo}
F=\frac{8}{16+\zeta^2+\zeta\sqrt{\zeta^2+16}} \; .
\eeq
The coherence visibility (\ref{jpjm}) for the state (\ref{gstwor}) is,
instead, given by
\beq
\label{jpjmtwo}
\alpha=\frac{4(\zeta+\sqrt{\zeta^2+16})}{16+\zeta^2+\zeta\sqrt{\zeta^{2}+16}}
\; .
\eeq
It is possible, also, to evaluate the entanglement entropy (\ref{ee})
when the system is in the state (\ref{gsetwor}).
We have that $S$ can be written as follows
\beqa
\label{eetwo}
S &=& -A^{2}\bigg(2\log_{2}\bigg[\frac{2
(\zeta+\sqrt{\zeta^2+16})^2}{\big(16+\zeta^2+\zeta\sqrt{\zeta^2+16}\big)^2}
\bigg]
\nonumber\\
&+&\big(\frac{\zeta^2+\zeta\sqrt{\zeta^2+16}}{4}\big)\log_{2}
\bigg[\frac{\big(\zeta+\sqrt{\zeta^2+16}\big)^2}{32+2\zeta^2+2\zeta
\sqrt{\zeta^2+16}}\bigg]\bigg)\;.\nonumber\\
\eeqa

As conclusive remarks, we observe that when $\zeta \rightarrow -\infty$
the state (\ref{gstwor})
tends to
\beq
\label{gstworcat}
|\psi_{0}\rangle=\frac{1}{\sqrt{2}}\bigg(|2,0\rangle+
|0,2\rangle\bigg) \;.\eeq
This is the macroscopic superposition state (\ref{cat}) with two bosons.
In the limit $\zeta \rightarrow -\infty$,
the energy (\ref{gstwor}) tends to $\displaystyle{U+\frac{4J^2}{U}}$.
The Fisher information (\ref{fitwo}) tends to $1$, the coherence
visibility (\ref{jpjmtwo}) to $0$, and the entanglement entropy
(\ref{eetwo}) to $1$.

When the limit $\zeta \rightarrow 0$ is met, the state (\ref{gstwor}) tends to
\beq
\label{gstworzu}
|\psi_{0}\rangle=\frac{1}{2}\bigg(|2,0\rangle+\sqrt{2}|1,1\rangle+
|0,2\rangle\bigg)\nonumber\\ \; ,\eeq
which is the atomic coherent state (\ref{coherent}) with two bosons.
In this case, the energy (\ref{gsetwor}) tends to $-2J$.
%which is the so-called bandwidth.
In the above mentioned limit, the Fisher information (\ref{fitwo})
tends to $1/2$, the coherence visibility (\ref{jpjmtwo}) to $1$, and the
entanglement entropy (\ref{eetwo}) to $3/2$, in agreement with the
first row of Tab. 1.

Finally, let us analyze the limit $\zeta \rightarrow +\infty$. We have that the
ground-state (\ref{gstwor}) tends to
\beq
\label{gstworfock}
|\psi_{0}\rangle=|1,1\rangle
\;,\eeq
which is the separable Fock state (\ref{Fock}) with two bosons.
In the aforementioned limit,
%the energy (\ref{gsetwor}) tends to $-4 J^2/U$
the Fisher information (\ref{fitwo}),
the coherence visibility (\ref{jpjmtwo}), and the entanglement entropy
(\ref{eetwo}) tend to $0$.

\subsection{Quasi-classical coherent state with number fluctuations}

In this subsection, for the sake of completeness, 
we shall compare the $N$-boson ground-state
with the quasi-classical coherent state as follows. 

The familiar Josephson equations \cite{smerzi,leggett}
of the bosonic Josephson junction can be obtained from
the Bose-Hubbard Hamiltonian (\ref{twomode}) by using the
quasi-classical coherent state
$|QC\rangle$. This state is given by the tensor product
of the coherent state $|CS\rangle_{L}$, which describes
the Bose-Einstein condensate in the left well,
and the coherent state $|CS\rangle_{R}$,
which describes the Bose-Einstein condensate
in the right well. The state $|CS\rangle$ is thus given by
\beq
\label{CS}
|QC\rangle=|CS\rangle_{L} \otimes |CS\rangle_{R}
\;,\eeq
where $|CS\rangle_{k}$ ($k=L,R$) is such that \cite{glauber}
\beq
\label{CSI}
|CS\rangle_{k}=e^{-|z_k|^{2}/2}\sum_{n=0}^{\infty}\frac{z_{k}^n}
{\sqrt{n!}}|n\rangle \;.
\eeq
The complex quantity $z_k$ is the eigenvalue of the annihilator in the
$k$th well, i.e.
\beq
\label{zetai}
\hat{a}_k |CS\rangle_{k}=z_k\,|CS\rangle_{k} \; .
\eeq
The absolute values of the two $z_k$ are related to the average
occupancy of the two wells:
\beq
N_k\equiv \langle QC|{\hat n}_k|QC\rangle = |z_k|^2  \; .
\eeq
So $z_k$ are conveniently parametrized as
$z_k=\sqrt{N_k}\exp(i\theta_k)$, where $\theta_k$ are
phase variables. It is important to stress that
the quasi-classical coherent state $|QC\rangle$ does not
have a fixed number of particles
and consequently it allows for non vanishing
quantum fluctuations of the total number of bosons.
The average boson number in the state $|QC \rangle$ is simply given by
$N=N_L+N_R=|z_L|^2+|z_R|^2$.

\begin{figure}[ht]
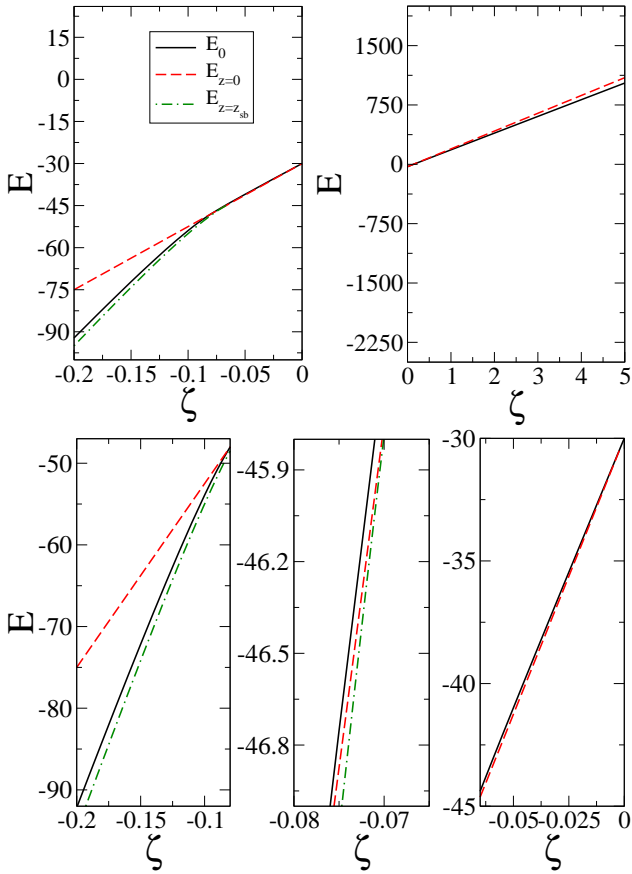

\epsfig{file=cats9-1.eps,width=0.96\linewidth,clip=}
\epsfig{file=cats9-2.eps,width=0.96\linewidth,clip=}
\caption{(Color online). On the vertical axis, the energy of the
ground-state $|E_0\rangle$,
 $E_0$ (solid line), the energy
(\ref{e}) calculated for $z=0$,  $E_{z=0}$ (dashed line), the energy
(\ref{e}) calculated for $z=z_{sb}$,  $E_{z=z_{sb}}$ (dot-dashed line). On the
horizontal axis the parameter $\zeta=U/J$. Left part of the upper panel: 
attractive bosons
($U<0$). Right part of the upper panel:
repulsive bosons ($U>0$). Lower panel: three insets for attractive bosons.
We have set $N=30$. $E$ is in units of $J$ and $\zeta$ is an adimensional
quantity.}
\label{fig6}
\end{figure}

The expectation value $E$ of the BH Hamiltonian
(\ref{twomode}) evaluated with respect to the state (\ref{CS}) is
\beq
\label{e}
E=\langle QC|{\hat H}|QC\rangle =
-N\,J\sqrt{1-z^2}\cos \phi +\frac{N^2\,U}{4}(1+z^2) \; ,
\eeq
where $z=(N_L-N_R)/N$ and $\phi=\theta_R-\theta_L$.
In the formula (\ref{e}) we recognize the total energy of a nonrigid
pendulum, which represents the mechanic analog of a bosonic Josephson junction
\cite{smerzi,leggett}. In fact, the state $|QC\rangle$ is expected to be
close to the ground state $|E_0\rangle$ in the weak-coupling regime
\cite{leggett}. It is straightforward to show that the
minima of the energy (\ref{e}) are obtained with $\phi=0$ and
\beqa
\label{minima}
&&z=z_{sb}=\pm \sqrt{1-\frac{4}{\zeta^2\,N^2}} \quad \mbox{if} \quad
\zeta<-\frac{2}{N}\nonumber\\
&&z=0 \quad \mbox{if} \quad \zeta>-\frac{2}{N}
\nonumber\\
\eeqa
Note that $\zeta=-2/N$ {\bf signals} the onset of the self-trapping regime
within an attractive 1D bosonic Josephson junction \cite{sb}.
The energy (\ref{e}) calculated for $z=0$, say $E_{z=0}$, and for $z=z_{sb}$,
say $E_{z=z_{sb}}$, becomes
\beqa
\label{em}
&&E_{z=0}=N\bigg(\frac{N\,U}{4}-1\bigg)
\\
&&E_{z=z_{sb}}=\frac{J^2}{U}+\frac{N^2\,U}{2}\; .
\eeqa
It is interesting to compare the energy $E$ of the quasi-classical
state $|QC\rangle$ with the energy $E_0$ of the state $|E_0\rangle$,
that is the ground-state of the BH Hamiltonian
with a fixed number $N$ of bosons (see also Ref. \cite{vittorio}).
>From the {\bf upper right} panel of Fig. \ref{fig6} - obtained with $N=30$
- we see that for repulsive bosons, $E_0$ (solid line) is smaller
than $E_{z=0}$ (dashed line). For attractive bosons, the situation is
much more complex.
Let us focus on the lower panels of  Fig. \ref{fig6}. We can see
that there are three different regions. In the first one (left panel:
$\zeta<-2/N-\Delta$ $(\Delta>0)$ ) we have that $E_{z=z_{sb}}<E_0<E_{z=0}$;
in the second
region (middle panel: $-2/N-\Delta<\zeta<-2/N$ ) we have that
$E_{z=z_{sb}}<E_{z=0}<E_{0}$; in the third region (right panel: $\zeta>-2/N$)
we have that $E_{z=0}<E_0$. Note that the greater is $N$ the smaller is
$\Delta$.

\begin{figure}[ht]
\epsfig{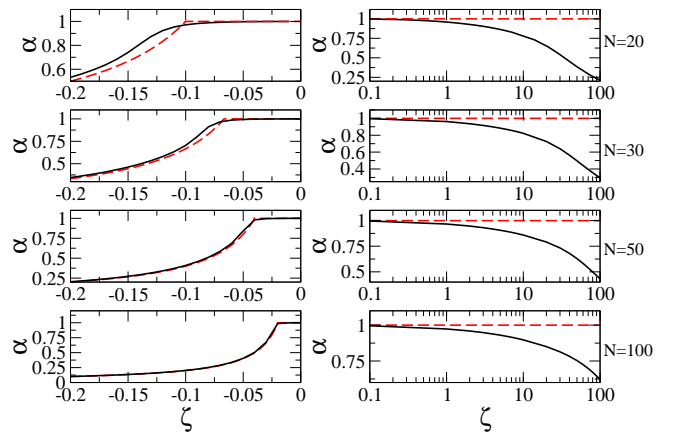}
\caption{(Color online). Coherence visibility $\alpha$ of the bosonic junction
as a function of the dimensionless parameter $\zeta = U/J$.
Left panel: attractive bosons ($U<0$). Right panel:
repulsive bosons ($U>0$). Solid line: results from Eq. (\ref{jpjm}). Dashed
line: results from Eq. (\ref{alphacs}).
Note that the horizontal axis of the right panel is in logarithmic scale.
The coherence visibility $\alpha$ and $\zeta$ are adimensional quantities.}
\label{fig7}
\end{figure}

The coherence visibility (\ref{alpha}) calculated with respect
to the state $|QC\rangle$ reads
\beq
\label{alphacs}
\alpha=\sqrt{1-z^2}
\;
\eeq
with $z$ given by Eqs. (\ref{minima}). In Fig. \ref{fig7} we report
the comparisons between the coherence
visibility calculated by using the ground-state $|E_0\rangle$
of the BH Hamiltonian
(\ref{twomode}), and that one
calculated by using Eq. (\ref{alphacs}).
Eq. (\ref{alphacs}) . Fig. \ref{fig7} clearly shows
that the coherence visibility $\alpha$ of the quasi-classical coherent
state $|QC\rangle$ is strictly equal to one only for a non-negative
interaction strength. Moreover, from the left panels (attractive
bosons) of Fig. \ref{fig7}, we can see that the
greater is the number of particles in the system the better is the agreement
between the predictions deriving from the two approaches.

The Fisher information $F$, Eqs. (\ref{qfigs})-(\ref{fi}), calculated
with respect
to the state $|QC\rangle$, is simply given by
\beq
\label{fcs}
F = {1\over N} \; .
\eeq
This is the same result that one obtains when the system
is in the atomic coherent state (\ref{coherent}), i.e. the exact
eigenstate for $U=0$.

\section{Conclusions}

We have investigated an atomic Josephson junction in the presence of
a finite number of interacting ultracold bosons confined in a 
quasi one-dimensional
geometry. By using the two-site Bose-Hubbard Hamiltonian toolbox, 
we have carried out the zero-temperature analysis by finding the
ground-state of the system. We have characterized the presence 
of macroscopic Schr\"odinger-cat states,
the coherence properties, and the quantum correlations by calculating 
the Fisher information, the visibility of the interference fringes
in the momentum distribution, and the entanglement 
entropy as functions of the boson-boson interaction strength.
The joint application of the Hellmann-Feynman theorem and 
of the perturbative approach
to evaluate the ground-state energy allows to obtain analytical
formulas - within the interactions range in which the 
perturbative theory is applicable
-  both for the Fisher information and for the coherence visibilty. 
The Fisher information and the
coherence visibility predicted by these formulas are in good
agreement with those evaluated by diagonalizing the 
two-site Bose-Hubbard Hamiltonian.
In the attractive 
regime we have found that the presence of macroscopic 
Schr\"odinger-cat states corresponds to a maximum of the Fisher 
information, while the entanglement
entropy has a maximum when the system begins to lose
coherence and to exhibit self-trapped dynamics. 
The maximal entanglement entropy is larger than that of a 
macroscopic cat state and in fact
very close to its theoretical upper limit.

GM thanks Alessio Serafini and S.M. Giampaolo for useful discussions.


\begin{thebibliography}{99}

\bibitem{oliver} O. Morsch and M. Oberthaler,
Rev. Mod. Phys. {\bf 78}, 179 (2006).

\bibitem{book-barone} A. Barone and G. Patern\`{o},
{\it Physics and Applications of the Josephson effect} (Wiley, New
York, 1982).

\bibitem{smerzi} S.Raghavan, A. Smerzi, S. Fantoni, R. Shenoy,
Phys. Rev. A {\bf 59}, 620 (1999).

\bibitem{stringa} L. Pitaevskii and S. Stringari,
Phys. Rev. Lett. {\bf 83}, 4237 (1999);
L. Pitaevskii and S. Stringari,
Phys. Rev. Lett. {\bf 87}, 180402 (2001).

\bibitem{anglin} J.R. Anglin, P. Drummond, and A. Smerzi,
Phys. Rev. A {\bf 64}, 063605 (2001).

\bibitem{mahmud} K.W. Mahmud, H. Perry, and W.P. Reinhardt,
J. Phys. B: At. Mol. Opt. Phys. {\bf 36}, L265 (2003);
K.W. Mahmud, H. Perry, and W.P. Reinhardt,
Phys. Rev. A {\bf 71}, 023615 (2005).

\bibitem{anna} G. Ferrini, A. Minguzzi, F. W. Hekking,
Phys. Rev. A {\bf 78}, 023606(R) (2008).

\bibitem{cirac} J.I. Cirac, M. Lewenstein, K. Molmer, and P. Zoller,
Phys. Rev. A {\bf 57}, 1208 (1998).

\bibitem{dalvit} D.A.R. Dalvit, J. Dziarmaga, and W.H. Zurek,
Phys. Rev. A {\bf 62}, 013607 (2000).


\bibitem{huang} Y.P. Huang and M.G. Moore,
Phys. Rev. A {\bf 73}, 023606 (2006).

\bibitem{carr} L.D. Carr, D.R. Dounas-Frazer, and
M.A. Garcia-March, EPL {\bf 90}, 10005 (2010).

\bibitem{brand} D.W. Hallwood, T. Ernst, and J. Brand,
e-preprint arXiv:1007.4038.

\bibitem{exp-bec} F.S. Cataliotti {\it et al.}, Science {\bf 293},
843 (2001); Y. Shin {\it et al.}, Phys. Rev. Lett. {\bf 92},
050405 (2004); M. Albiez {\it et al.}, {\it ibid.} {\bf 95},
010402 (2005); S. Levy {\it et al.}, Nature (London) {\bf 499},
579 (2007).

\bibitem{ajj1} G. Mazzarella, M. Moratti, L. Salasnich,
M. Salerno and F. Toigo,
J. Phys. B: At. Mol. Opt. Phys. {\bf 42}, 125301 (2009).

\bibitem{ajj2} G. Mazzarella, M. Moratti, L. Salasnich, F. Toigo,
J. Phys. B: Atom. Mol. Opt. Phys. {\bf 43}, 065303 (2010).

\bibitem{sb} G. Mazzarella and L. Salasnich, Phys. Rev. A  {\bf 82}, 033611
(2010).

\bibitem{leggett} A. J. Leggett, {\it Quantum Fluids} (Oxford University
Press, Oxford) (2006).

\bibitem{io-e-boris} L. Salasnich, B.A. Malomed, and F. Toigo,
Phys. Rev. A {\bf 81}, 045603 (2010).

\bibitem{oberthalernature} J. Est\`eve, C. Gross, A. Weller,
S. Giovanazzi, M. K.
Oberthaler, Nature {\bf 455}, 1216-1219 (2010); C. Gross, T. Zibold,
E. Nicklas, J.
Est\`eve, M. K. Oberthaler, Nature {\bf 464}, 1165-1169 (2010).

\bibitem{oberthalerpra10} C. Bodet, J. Est\`eve, M. K. Oberthaler, T. Gasenzer,
Phys. Rev. A {\bf 81}, 063605 (2010).

\bibitem{cohen} C. Cohen-Tannoudji, B. Diu, F. Laloe,
{\it Quantum Mechanics}, Vol. 2, (J. Wiley, New York, 1977).

\bibitem{pezze} L. Pezz\`{e} and A. Smerzi,
Phys. Rev. Lett. {\bf 102}, 100401 (2009).

\bibitem{fermidyn} L. Salasnich, G. Mazzarella, M. Salerno, and
F. Toigo, Phys. Rev. A {\bf 81}, 023614 (2010).

\bibitem{arecchi} F.T. Arecchi, E. Courtens, R. Gilmore, H. Thomas, Phys. Rev.
A {\bf 6}, 2211 (1972); G. J. Milburn, J. Corney, E. M. Wright, D. F. Walls,
Phys. Rev. A {\bf 55}, 4318 (1997).

\bibitem{diaz} B. Julia-Diaz, D. Dagnino,
M. Lewenstein, J. Martorell, and A. Polls, Phys. Rev. A {\bf 81},
023615 (2010).

\bibitem{landau} L. Landau and L. Lifshitz,
{\it Course in Theoretical Physics}, Vol. 3, {\it Quantum
Mechanics: Non-Relativistic Theory}, (Pergamon, New York, 1959).

\bibitem{braunstein} W.K. Wootters, Phys. Rev. D. {\bf 23}, 357 (1981); C. W.
Helstrom, {\it Quantum Detection and Estimation Theory} (Academic Press,
New York, 1976), Chap. VIII; A. S. Holevo, {\it Probalistic and Statistical
Aspect of Quantum Theory} (North-Holland, Amsterdam, 1982); 
S.L. Braunstein and C. M. Caves,
Phys. Rev. Lett. {\bf 72}, 3439 (1994).

\bibitem{weiss} B. Gertjerenken, S. Arlinghaus, N. Teichmann, C.Weiss
Phys. Rev. A {\bf 82}, 023620 (2010).

\bibitem{note}
We note that for $N=2$, the second order perturbation theory 
lifts the degeneracy 
between the even and the odd parity states, which are shifted by
$\pm 2J^2/U$ with  respect to the classical value (see Eq. \ref{gsetwor}) 
leading to a splitting
$\displaystyle{\Delta E =\frac{4J^2}{|U|}}$.
For $N>2$ second order perturbation theory does not remove the ground
state degeneracy.

\bibitem{bwae} C.H. Bennett, H.J. Bernstein, S. Popescu, B. Schumacher,
Phys. Rev. A {\bf 53}, 2046 (1996); S. Hill and W. Wootters,
Phys. Rev. Lett. {\bf 78}, 5022 (1997); L. Amico, R. Fazio,
A. Osterloh, V. Vedral, Rev. Mod. Phys.
{\bf 80}, 517 (2008); J. Eisert, M. Cramer, M. B. Plenio, Rev. Mod. Phys.
{\bf 82}, 277 (2010).

\bibitem{glauber} R. J. Glauber, Phys. Rev. {\bf 131}, 2766 (1963).

\bibitem{vittorio} P. Buonsante, P. Kevrekidis, V. Penna, and A. Vezzani,
J. Phys. B: At. Mol. Opt. Phys. {\bf 39}, S77 (2006).

\end{thebibliography}
\end{document}